\documentclass[twocolumn,english,notitlepage,superscriptaddress]{revtex4-1}
\usepackage[LGR,T1]{fontenc}
\usepackage[latin9]{inputenc}
\usepackage{color}
\usepackage{amsmath}
\usepackage{amssymb}
\usepackage{bm}
\usepackage{graphicx}
\usepackage{wasysym}

\usepackage{graphicx}

\makeatletter

\DeclareRobustCommand{\greektext}{%
  \fontencoding{LGR}\selectfont\def\encodingdefault{LGR}}
\DeclareRobustCommand{\textgreek}[1]{\leavevmode{\greektext #1}}
\ProvideTextCommand{\~}{LGR}[1]{\char126#1}

\usepackage{comment}
\newcommand{\prlsec}[1]{\textit{#1.---}}

\usepackage{babel}

\makeatother

\usepackage{babel}
\begin{document}
\title{Creating atom-nanoparticle quantum superpositions}
\author{M. Toro\v{s}}
\affiliation{School of Physics and Astronomy, University of Glasgow, Glasgow, G12
8QQ, UK}
\affiliation{Department of Physics and Astronomy, University College London, Gower
Street, WC1E 6BT London, UK}
\author{S. Bose}
%\email{s.bose@ucl.ac.uk}

\affiliation{Department of Physics and Astronomy, University College London, Gower
Street, WC1E 6BT London, UK}
\author{P. F. Barker}
%\email{p.barker@ucl.ac.uk}

\affiliation{Department of Physics and Astronomy, University College London, Gower
Street, WC1E 6BT London, UK}
\begin{abstract}
A nanoscale object evidenced in a non-classical state of its centre
of mass will hugely extend the boundaries of quantum mechanics. To
obtain a practical scheme for the same, we exploit a hitherto unexplored
coupled system: an atom and a nanoparticle coupled by an optical field.
We show how to control the center-of-mass of a large $\sim500$nm
nanoparticle using the internal state of the atom so as to create,
as well as detect, nonclassical motional states of the nanoparticle.
Specifically, we consider a setup based on a silica nanoparticle coupled
to a Cesium atom and discuss a protocol for preparing and verifying
a Schrödinger-cat state of the nanoparticle that does no require cooling
to the motional ground state. We show that the existence of the superposition
can be revealed using the Earth's gravitational field using a method
that is insensitive to the most common sources of decoherence and
works for any initial state of the nanoparticle. 
\end{abstract}
\maketitle
\prlsec{Introduction} Quantum mechanics has been probed experimentally
over a vast range of energies and scales. On the one side, down to
subatomic distances using accelerators, while on the other side, spatial
superpositions in the mesoscopic regime are being explored via quantum
optomechanics. The former is ultimately expected to shed light on
the basic building blocks of our universe, while the latter addresses
the quantum-to-classical transition in the mesoscopic, a problem already
highlighted by Schrödinger~\citep{schrodinger1935gegenwartige}.

The field of optomechanics, and in particular levitated optomechanics~\citep{Millen_2020},
where the system is well isolated from deleterious effects of decoherence
from the environment, has now reached the quantum regime~\citep{delic2019motional,tebbenjohanns2020motional}
and is expected to soon test ideas from quantum foundations~\citep{bassi2013models}
and the nature of gravity~\citep{bose2017spin,marletto2017gravitationally,marshman2019locality}.
Nonetheless, a challenge still remains how to prepare nonclassical
motional states of the nanoparticle, such as the Schrödinger-cat state~\citep{hacker2019deterministic}.

Possible approaches for nonclassical state preparation in levitated
optomechanics are based on nonlinearities in the potential~\citep{ralph2018dynamical},
as well as coupling to quantized fields along with possible usage
of measurements~\citep{bose1997preparation,bose1999scheme,romero2011large,vanner2011selective,brawley2016nonlinear,clarke2018growing}.
Difficulties of these approaches include small single photon nonlinearities
and/or detecting the effect of nonlinerities in the regime of small
oscillations, where the motion is typically well described by a linear
theory. Another promising strategy is to embed impurities in the nanoparticle
and use that to control the nano-particle~\citep{kolkowitz2012coherent,arcizet2011single,scala2013matter,yin2013large,wan2016free}.
However, the placement, control and coherence of such impurities is
experimentally very challenging. Hence any alternatives which are
not susceptible to the above limitations are highly desirable.

Here we propose combining two hitherto disparate fields in an optimal
way for the nonclassical state preparation of nano-objects: the long
acquired ability to control the exceptionally coherent internal levels
of trapped atoms (ions), and through them, their motional states~\citep{monroe1996schrodinger}
and the recently acquired expertise of controlling, to an exceptional
level, the centre of mass of nano-objects~\citep{delic2019motional,tebbenjohanns2020motional}.
We show how the addition of the highly controllable atom opens up
feasible opportunities for the preparation of Schrödinger Cat states
in the latter field. We consider the situation where the nanoparticle
is trapped in a Paul trap and illuminated by a plane-wave optical
field. The reflected light from the nanoparticle interferes with the
incoming light and creates a series of dipole traps where atoms can
be trapped. In particular, we consider one atom placed in a stiff
trap such that displacing it also moves the center-of-mass of the
atom-nanoparticle system. The induced effective coupling between the
motional state of the nanoparticle and the internal state of the atom
allows to directly apply the technical abilities from atomic physics
to prepare non-classical states of the nano-object. Moreover, the
switchability of the coupling (simply by controlling the intensity
of the optical field) enables release and recapture so as to exploit
free-fall non-decoherent evolutions. This latter ability, for example,
is absent in atom-micromechanical coupled systems~\citep{hammerer2010optical,vogell2013cavity,bennett2014coherent,ranjit2015cold}.
We show that one can generate a small spatial superposition of the
nanoparticle so that it is well protected from enviromental decoherence,
and yet such a small superposition can be revealed using the Earth's
gravitational field~\citep{scala2013matter,rademacher2020quantum}.
Moreover, we find that the protocol is insensitive to the initial
state of the nanoparticle which will greatly facilitate the realization.

\begin{figure}[t]
\includegraphics[width=1\columnwidth]{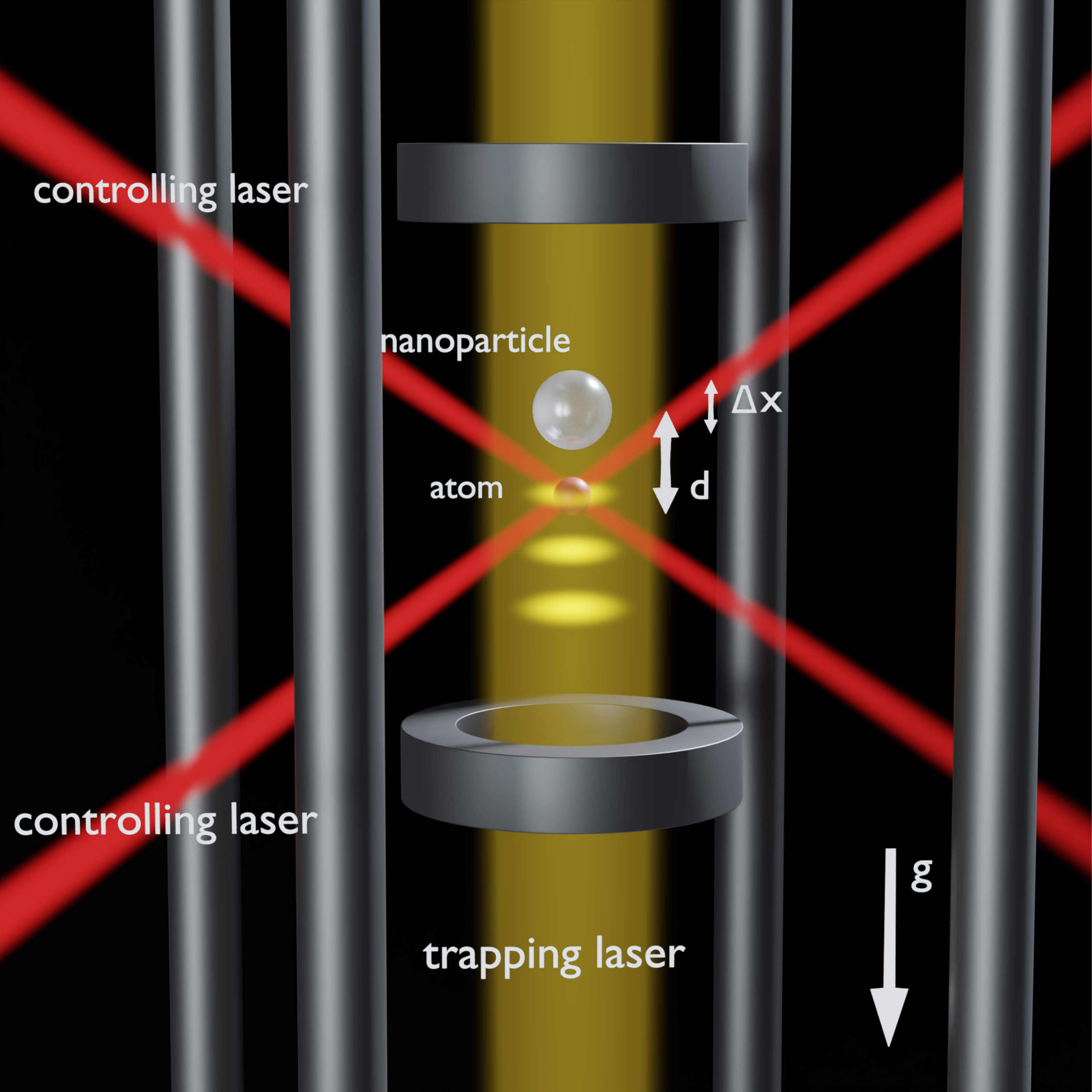} \caption{Scheme of the experimental setup. A nanoparticle of mass $m_{n}$
is trapped in the Paul trap. A plane wave optical field illuminates
the nanoparticle and the backscatter interferes to create an intnesity
maxima at distance $d$ below the nanoparticle, where we trap an atom.
For a very stiff atomic trap we obtain an effective coupling between
the internal state of the atom and the nanoparticle. The initial height
of the nanoparticle in the trap can be controlled by changing the
power of the trapping laser, which can be switched off quickly, together
with softening the Paul trap to very low frequencies, approximately
obtaining a free fall regime for a time $\Delta t$. We create and
control the spatial superposition of the nanoparticle using additional
lasers coupled to hyperfine transitions, labelled as controlling lasers,
with the superposition size denoted by $\Delta x$. At the end we
perform a readout of the accumulated gravitational phase $\phi_{\text{grav}}\sim m_{n}g\Delta x\Delta t/\hbar$
using a cycling transition of the internal state, where $g$ is the
Earth's gravitational acceleration.}
\label{setup} 
\end{figure}

\begin{figure}
\includegraphics[width=1\columnwidth]{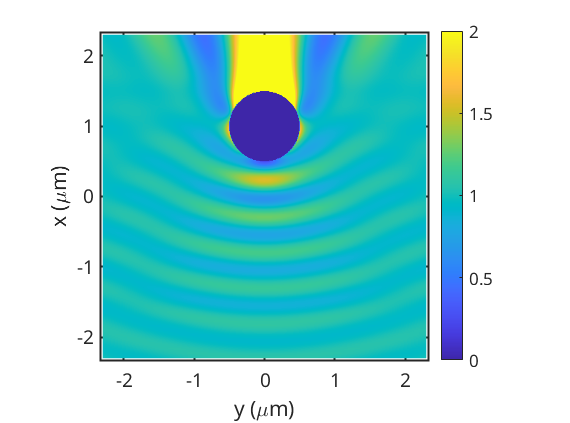}

\caption{Simulated intensity using finite difference time domain methods~\citep{sim1,lumerical}.
We consider a nanoparticle of radius $r=500$nm and an optical field
with wavelength $\lambda_{l}=1000$nm propagating in the posive $x$-axis
direction. The incoming field is polarized along the $y$-axis; other
vertical planes shows a similar intensity profile. The colour bar
is the enhancement in the square of the electric field. The large
blue circle denotes the nanoparticle; the incoming field propagating
from the bottom interferes with the backcattered field from the nanoparticle
which creates dipole traps below the nanoparticle. {The
strongest dipole trap is located $d\sim0.75\text{\ensuremath{\mu}m}$
below the center of the nanoparticle (first yellow patch below the
blue circle). }}

\label{simulation} 
\end{figure}

\prlsec{Atom-nanoparticle coupling} The experimental setup consists
of a nanoparticle trapped in Paul trap which is illuminated by a plane-wave
optical field (see Fig.~\ref{setup}). We choose the light wavelength
$\lambda_{l}$ to be comparable or smaller than the nanoparticle radius
$r$, effectively making the nanoparticle a mirror-like object. The
backscattered light from the nanoparticle interferes with the incoming
light to form a standing wave in the rest frame of the nanoparticle
(see Fig.~\ref{simulation}) and the resulting intensity minima and
maxima rigidly follow the motion of the nanoparticle. In one of the
maxima we trap an atom exploiting an internal electronic transition
in the red-detuned regime. Specifically, the potential is given by:
\begin{equation}
\hat{H}_{\text{trap}}=\frac{m_{n}\omega_{n}^{2}}{2}\hat{x}_{n}^{2}+\frac{m_{a}\omega_{a}^{2}}{2}\left[\hat{x}_{a}-(\hat{x}_{n}+d)\right]^{2},\label{eq:trap}
\end{equation}
where $\omega_{n}$ ($\omega_{a}$) is the frequency of the Paul (atomic)
trap, $m_{n}$ ($m_{a}$) is the mass of the nanoparticle (atom),
$\hat{x}_{n}$ ($\hat{x}_{a}$) is the nanoparticle (atom) position,
and $d$ is the distance between the two traps.

The motional frequency of the atom is given by~\citep{grimm2000optical}:
\begin{equation}
\omega_{a}=\sqrt{\frac{6\pi c^{2}}{m_{a}w^{2}\text{\ensuremath{\omega_{e}^{3}}}}I\frac{\Gamma}{\Delta}},
\end{equation}
where $I$ is the intensity of light at the trap center, $w\sim\lambda_{l}/2$
is the trap width, $\omega_{e}$ is the electronic transition frequency,
$\Gamma$ is the decay rate from the excited state, $\Delta=\omega_{e}-\omega_{l}$
is the detuning of the light field, $\omega_{l}=\frac{2\pi c}{\lambda_{l}}$,
and $c$ is the speed of light. To obtain high trapping frequencies
we can decrease the detuning $\Delta$ at the cost of reducing the
trapping time $\tau_{\text{trap}}=\frac{m_{a}c^{2}}{\hbar\text{\ensuremath{\omega_{l}}}^{2}}\frac{\Delta}{\Gamma}$.

The trapped atom offers a new handle on motion of the nanoparticle.
Particularity interesting is the situation when the atom is placed
in a strong dipole trap, resulting in a rigid atom-nanoparticle coupling.
We then expect that any displacement of the atom will drag the whole
atom-nanoparticle system, with only negligible excitation of the relative
motion between the two. Mathematically, this translates to requiring
that (i) the atom is placed in the motional ground state and (ii)
the zero-point motion of the atom, $\delta_{a}$, is small with respect
to the one of the nanoparticle, $\delta_{n}$, such that when the
nanoparticle is excited the atom remains in the ground state, i.e.
we can write $\hat{x}_{a}\approx\hat{x}_{n}-d$.

\prlsec{Nanoparticle motion control} In the considered regime we
find the following interaction Hamiltonian between the \emph{motional
state of the nanoparticle} and \emph{the atomic hyperfine transition}
(in interaction picture) 
\begin{alignat}{1}
\frac{\hat{H}_{\text{int}}}{\hbar}= & \frac{\Omega_{jk}}{2}\sigma_{+}\:\text{exp}\left(i\left[\eta(\hat{a}e^{-i\omega_{n}t}+\hat{a}^{\dagger}e^{-i\omega_{n}t})-\delta t+\phi\right]\right)\nonumber \\
 & +\text{H.c},\label{eq:interaction}
\end{alignat}
where we have introduced the nanoparticle mode $\hat{a}$, i.e. $\hat{x}_{n}=\delta_{n}(\hat{a}^{\dagger}+\hat{a})$.
$\Omega_{jk}$ is the coupling of the stimulated Raman transition
between the hyperfine states $\vert j\rangle$ and $\vert k\rangle$,
$\sigma_{+}=\vert k\rangle\langle j\vert$, $\eta=k\delta_{n}$ is
the Lamb-Dicke parameter, $k=\frac{2\pi}{\lambda}=\frac{\omega}{c}$
with $\omega$ the frequency of the laser, $\delta=\omega_{h}-\omega$
is the detuning that selects one of the sidebands or the carrier resonance,
$\omega_{h}$ is the hyperfine transition frequency, and $\phi$ is
a phase that includes $\frac{d}{\lambda}$. Here we limit the discussion
to $\eta\ll1$, which puts a lower bound on the Paul trap frequency,
i.e. $\frac{\hbar}{2m_{n}\lambda^{2}}\ll\omega_{n}$. The coupling
of the stimulated Raman transition is given by $\Omega_{jk}\equiv g_{jk}$,
where $g_{jk}=\frac{qE}{\hbar}D_{jk},$ $q$ is the electron charge,
$E$ is the amplitude of the electric field, and $D_{jk}$ is the
transition dipole matrix element between the state $j$ and $k$.

We are interested in two types of interactions, one that (a) controls
the internal state without affecting the motional state, and one that
(b) displaces the motional state without changing the internal one,
both of which can be implemented in a $\Lambda$-type scheme using
two lasers. In particular, using two-photon stimulated Raman transitions
of type (a) and (b) we will consider three types of operations, where
the coupling will be given by $\Omega_{jk}\equiv\frac{g_{jl}^{*}g_{lk}}{\Delta_{l}}$,
and $\Delta_{l}$ is the detuning from the intermediate state $l$~\citep{wineland1998experimental}.
To create a superposition of the hyperfine states we consider the
carrier frequency, i.e. $\delta=0$, with a pulse of duration $t=\pi/(2\Omega_{\uparrow\downarrow})$
using scheme (a), namely a $\pi/2$ pulse. This generates a beam splitter
transformation, i.e. the hyperfine states evolve in the following
way: $\vert\uparrow\rangle\rightarrow(\vert\uparrow\rangle-\vert\downarrow\rangle)/\sqrt{2}$
and $\vert\downarrow\rangle\rightarrow(\vert\uparrow\rangle+\vert\downarrow\rangle)/\sqrt{2}$.
Similarly, a $\pi$ pulse using scheme (a) at the carrier corresponds
to $\Omega_{\uparrow\downarrow}t=\pi$ and $\delta=0$, which exchanges
the hyperfine states, i.e. $\vert\uparrow\rangle\rightarrow-\vert\downarrow\rangle$
and $\vert\downarrow\rangle\rightarrow\vert\uparrow\rangle$. On the
other hand, to displace the motional state without modifying the hyperfine
state we exploit scheme (b) at the first red sideband, i.e $\delta=\omega_{n}$.
This latter operation produces a displacement of the motional state
by $\Omega_{\downarrow\downarrow}\eta t$, where $t$ is the duration
of the pulse.

In summary, the discussed interactions have the same form as the ones
exploited in atomic physics where in place of the motional state of
the atom we have the motional state of the nanoparticle. We can thus
adopt the experimentally well-established protocols from atomic physics
to the nanoscale~\citep{monroe1996schrodinger,itano1997quantum,wineland1998experimental}.

\prlsec{Schrödinger's cat} Suppose the state of the system is $\vert\Psi\rangle=\vert\psi\rangle_{h}\vert\psi\rangle_{n}$,
where $\vert\psi_{h}\rangle$ is the hyperfine state of the atom,
and $\vert\psi\rangle_{n}$ is the motional state of the nanoparticle.
Ideally, one would like to prepare a state of the form $\vert\psi\rangle_{n}\sim\vert\downarrow\rangle_{h}\vert\alpha_{\text{\text{top}}}\rangle_{n}+\vert\uparrow\rangle_{h}\vert\alpha_{\text{bottom}}\rangle_{n},$
where $\vert\alpha_{\text{\text{top}}}\rangle_{n}$ and $\vert\alpha_{\text{bottom}}\rangle_{n}$
denote states located at different heights in the Paul trap, i.e.
a Schrödinger-cat state. Once such a state has been created we then
want to ascertain its existence using as the readout the hyperfine
state $\vert\psi_{h}\rangle$.

A possible strategy is to cool the system to the ground state, i.e.
$\vert\Psi_{\text{init}}\rangle=\vert\downarrow\rangle_{h}\vert0\rangle_{n}$,
and to apply the procedure described by Monroe et al \citep{monroe1996schrodinger},
which consists of $\pi/2$, $\pi$, and displacement pulses. To make
such a scheme work one would however need additional optical fields
to control the motional state of the nanoparticle. In particular,
cooling to the motional ground state can be achieved with a cavity-tweezer
setup~\citep{delic2019motional} and is expected to be soon available
also in a tweezer setup~\citep{hebestreit2018sensing,tebbenjohanns2020motional}.

However, a protocol that would not require cooling~\citep{ranjit2015cold},
but would rather work for a generic trapped state, such as the experimentally
more readily available thermal state, is still desirable. A second
attractive feature would be to have a reliable method to evidence
that the nanoscale superposition has really been probed, for example,
by relating the outcome of the experiment to one of its intrinsic
properties such as the nanoparticle mass $m_{n}$. A possible strategy
to address both of these requirements has been outlined in \citep{scala2013matter},
parts of which we now adapt to the hybrid atom-nanoparticle system.
For simplicity of presentation we first consider the initial state
$\vert\psi_{\text{init}}\rangle=\vert\alpha\rangle\otimes\vert\downarrow\rangle$,
where the nanoparticle is prepared in the coherent state $\vert\alpha\rangle$
{(but we show below that it applies for }{\emph{any}}{{}
initial state)}. The protocol consists of the following steps. 
\begin{enumerate}
\item Trap a nanoparticle in the Paul trap at frequency $\omega_{1}$. Trap
an atom in an intensity maxima below the nanoparticle using a plane
wave and cool it to the ground state using resolved sideband cooling~\citep{wineland1998experimental}. 
\item Apply a $\pi/2$ pulse to generate the state $\vert\psi\rangle\sim\vert\alpha\rangle\otimes\left(\vert\downarrow\rangle+\vert\uparrow\rangle\right)$. 
\item Soften the Paul trap to frequency $\omega_{n}=\omega_{2}\ll\omega_{1}$. 
\item Apply a displacement beam for a time $\delta t$ to produce the state
$\vert\psi\rangle\sim\left(\vert\alpha+\beta\rangle\otimes\vert\downarrow\rangle+\vert\alpha\rangle\otimes\vert\uparrow\rangle\right)$,
where $\beta=\Omega_{gg}\eta\delta t$. \label{enu:Soften-the-Paul} 
\item Reduce the trapping laser power such that the radiation pressure force
becomes small and the nanoparticle-atom system starts falling towards
the Earth{{} (matter-wave coherence is thus shielded
from the deleterious effects of the laser photons and the system becomes
a matter-wave sensor for the local Earth's gravitational acceleration
$\sim g$).}
\item Leave the system in free fall for a time $\Delta t$ such the gravitational
field induces the phase $\phi_{\text{grav}}$: $\vert\psi\rangle\sim\left(e^{-i\phi_{\text{grav}}}\vert\alpha'+\beta\rangle\otimes\vert\downarrow\rangle+\vert\alpha'\rangle\otimes\vert\uparrow\rangle\right)$,
where $\vert\alpha'\rangle$ is the time-evolved coherent state of
$\vert\alpha\rangle$ . 
\item Increase the trapping laser power back to its initial value. Apply
a displacement beam for a time $\delta t$ to reverse the effect of
step \ref{enu:Soften-the-Paul} and obtain a factorizable state $\vert\psi\rangle\sim\vert\alpha'\rangle\otimes\left(e^{-i\phi_{\text{grav}}}\vert\downarrow\rangle+\vert\uparrow\rangle\right)$. 
\item Apply a $\pi/2$ pulse to create the final state $\vert\psi\rangle\sim\vert\alpha'\rangle\otimes\vert\phi\rangle$,
where the hyperfine state is $\vert\phi\rangle=\text{cos }\left(\frac{\phi_{\text{grav}}}{2}\right)\vert\downarrow\rangle-\text{sin}\left(\frac{\phi_{\text{grav}}}{2}\right)\vert\uparrow\rangle.$ 
\item Apply a laser field to drive a cycling transition and find the probability
of being in the ground state $P_{\downarrow}=\text{cos}^{2}\left(\frac{\phi_{\text{grav}}}{2}\right)$. 
\item After the measurement we recapture the nanoparticle by modulating
the radiation pressure from the trapping laser and the Paul trap frequency. 
\end{enumerate}
The induced gravitational phase difference is given by 
\begin{equation}
\phi_{\text{grav}}=\frac{m_{n}g\Delta x\Delta t}{\hbar},\label{eq:gphase}
\end{equation}
where $\Delta x=\delta_{n}\beta=\frac{\hbar k}{2m_{n}\omega_{2}}\Omega_{gg}\delta t$
is the superposition size of the nanoparticle and $\Delta t$ is the
duration of the transient free fall motion. Since the nanoparticle
mass $m_{n}$ is large we can have $\phi_{\text{grav}}\sim1$ already
for small superposition sizes $\Delta x$ and for short free-fall
times $\Delta t$ -- a regime which is interesting on its own.

Let us now consider a generic initial state $\rho_{\text{init}}=\rho_{n}\otimes\vert\downarrow\rangle\langle\downarrow\vert$,
where $\rho_{\text{\text{n}}}=\int d^{2}\alpha P_{n}(\alpha)\vert\alpha\rangle\langle\alpha\vert,$
and $P$ is Glauber's P quasi-probability distribution. Here we only
require that the nanoparticle is initially trapped in the Paul trap,
but the motional state can be otherwise completely generic. The steps
1-7 now result in the final state $\rho_{\text{final}}\sim\rho'_{n}\otimes\vert\phi\rangle\langle\phi\vert$,
where $\rho'_{n}$ is the final motional state of the nanoparticle,
yet $\vert\phi\rangle$ is the same internal state obtained by considering
an initial coherent motional state. Remarkably, the transient free
fall dynamics entangles the motional and internal states in a simple
way which can be readily disentangled at any time --- this is a direct
consequence of the uniform nature of the universal gravitational coupling,
a feature which is absent already with a harmonic potential. Creating
a superposition of an arbitrary motional state {(such
as of a thermal state)} still fully retains its coherent properties,
and once the gravitational phase is transferred to the internal state
it can be then read out again using steps 8 an 9.

\prlsec{Discussion}{{} }We can estimate the requirements
to achieve $\phi_{\text{grav}}\sim1$ for a typical tabletop experiment
using a nanoparticle of radius $r=500\text{nm}$ and mass $m_{n}\sim10^{-15}\text{kg}$
in a Paul trap~\citep{bullier2020characterisation,pontin2019ultranarrow}.
As discussed, we first trap an atom in a dipole trap near the nanoparticle,
which induces a coupling between the two, while other interactions
between the atom and the charged nanoparticle are negligible. For
concreteness we consider a Cs atom and the $D_{2}$ transition $6^{2}S_{\frac{1}{2}}\rightarrow6^{2}P_{\frac{3}{2}}$
which has a transition dipole matrix element $\sim4\times10^{-29}\text{Cm}$
and decay rate $\Gamma\sim3\times10^{7}\text{Hz}$.

We set the detuning of the trapping laser to $\Delta\sim5\times10^{11}\text{Hz}$
to generate a far red-detuned dipole trap: we find a trap lifetime
$\tau_{\text{trap}}\sim1\text{s}\gg\Delta t$ and using Fig.~\ref{simulation}
we estimate the atomic trap frequency to be $\omega_{a}\sim5\times10^{6}\text{Hz}$
{generated by an incoming (backscattered) intensity
$\sim5\times10^{12}\text{Wm}^{-2}$($\sim3\times10^{7}\text{Wm}^{-2}$)}.
Such an intensity can be obtained using an unfocused laser beam at
moderate power; at this intensity the radiation pressure force cancels
the gravitational one{{} (whilst not co-trapping the
nanoparticle}). We consider a short free fall-time $\Delta t\sim\omega_{a}^{-1}\sim1\text{\ensuremath{\mu}s}$
in order to retain the atom's motional state which corresponds to
a displacement of $\sim5\text{pm}$. The condition to excite the nanoparticle
motion constrains the Paul trap frequency $\omega_{n}$ from above,
$\omega_{n}\ll5\times10^{-4}\text{Hz}$, and the Lamb-Dicke condition
from below, $\omega_{n}\gg5\times10^{-8}\text{Hz}$. Specifically,
we set the initial Paul trap frequency to $\omega_{1}=0.1\text{kHz}$
which is then softened to $\omega_{2}=5\times10^{-6}\text{Hz}$. After
the Paul trap is softened we create a spatial superposition of the
nanoparticle by illuminating the atom with a short laser pulse of
duration $\sim100\text{ps}$ and detuning $\Delta_{3}\sim10^{11}\text{Hz}$.
The requirement of unit phase, $\phi_{\text{grav}}\sim1$, fixes the
intensity of the beam to $I\sim1\text{Wm}^{-2}$, resulting in a tiny
nanoparticle superposition of size $\Delta x\sim10^{-14}\text{m}$.
{The control beam will illuminate also the nanoparticle
(given its close proximity $d\sim0.75\mu m$), but such a tiny intensity
will however not lead to any measurable dephasing.} Larger as well
as smaller superpositions can be created by varying the parameters
of the setup, for example, by controlling the intensity and duration
of the displacement beam one is expected to achieve superpositions
of the size of the nanoparticle. Additionally, to further enlarge
the size of the superposition --without extending the duration of
the experiment -- one could also introduce a boosting potential by
adaptation of the coherent inflation method to the Paul trap~\citep{romero2017coherent}.

{The decoherence times for superposition sizes $\Delta x\sim10^{-14}\text{m}$
exceed the duration of the experimental time $\Delta t\sim1\mu s$
at readily available pressures and temperatures -- for concreteness
we consider the vacuum chamber with pressure $p\sim10^{-2}\text{mbar}$
and temperature $T\sim300\text{K}$. Given the modest laser intensities,
and the relatively high pressure, we can assume that both the center-of-mass
and internal temperature of the nanoparticle remain below $T\sim1000\text{K}$
\citep{hebestreit2018measuring} (for cooling the internal temperature
see \citep{rahman2017laser}). At such pressures/temperatures we find
that gas collisions limit the coherence time to $\sim6\mu s$, while
decoherence due to photon emission/absorption remains negligible --
at $T\sim300\text{K}$ the available coherence time is further extended~\citep{schlosshauer2007decoherence,romero2011quantum,seberson2019distribution}.}

{For completeness we also estimate the emitted thermal
radiation from the nanoparticle and its effect on the atom. Assuming
black-body radiation from the nanoparticle with internal temperature
$T\sim1000\text{K}$ we find a radiated intensity $\sim10^{5}\text{Wm}^{-2}$
which is two orders below the intensity generating the atom's dipole
trap (see above). Furthermore, the intensity of the thermal radiation
in the narrow frequency range of the internal transition Cs $D_{2}$($6^{2}S_{\frac{1}{2}}\rightarrow6^{2}P_{\frac{3}{2}}$)
is $\sim10^{-6}\text{Wm}^{-2}$ which has to be compared with the
intensity of the controlling lasers $\sim1\text{Wm}^{-2}$. We have
to however re-scale the two intensities by the ratio of the duration
of the experiment ($\sim1\text{\textgreek{m}s}$ and of the controlling
pulse and $\sim100\text{ps}$) which nonetheless still results in
the coherent laser radiation dominating by 2 orders of magnitude over
the thermal one. If instead one assumes an internal temperature $T\sim300\text{K}$
the effect of thermal radiation becomes dwarfed by the controlling
beams by about $\sim20$ orders of magnitude and can thus be again
neglected.}

{Finally, we estimate the effect of voltage noise,
$S_{V}$, which gives rise to a force noise, $S_{f}^{\text{(vol)}}\sim qS_{V}/D$,
where $q$ is the net charge on the nanoparticle, and $D$ is a characteristic
distance to the electrodes. Specifically, assuming $S_{V}\sim10\text{\ensuremath{\mu}V}/\text{Hz}^{1/2}$,
$q\sim80e$ (we note that the charge on the nanoparticle can be controlled
to a high degree \citep{bullier2020characterisation}), and $D\sim\text{2.3\text{mm}}$
we find $S_{f}^{\text{(vol)}}\sim10^{-23}\text{N}/\text{Hz}^{1/2}$\citep{pontin2019ultranarrow}.
By comparison the force noise due to gas collisions is $S_{f}^{\text{(gas)}}\sim\sqrt{2k_{b}Tm_{n}\gamma}$,
where $\gamma=4\pi m_{g}r^{2}v_{t}p/(3k_{b}Tm_{n})(1+\pi/8)$ is the
gas damping rate~\citep{epstein1924resistance,cavalleri2010gas},
$m_{g}$ is the molecular mass, and $v_{t}=\sqrt{8k_{b}T/(\pi m_{n})}$
is the thermal gas velocity -- using $T\sim300\text{K}$ and $p\sim10^{-2}\text{mbar}$
we find $S_{f}^{\text{(gas)}}\sim10^{-16}\text{N}/\text{Hz}^{1/2}$.
As discussed above the thermal noise does not impede the witnessing
of interference and hence voltage noise can be also safely neglected.}

{The insensitivity of the ten-step protocol to the
environment can be explained by the fact that the characteristic wavelength
of gas particles as well as the ones associated with laser and environmental
photons, is much larger than $\Delta x$, making the associated decoherence
times long compared to the short free fall time. }

In summary, we have shown that it is possible to create motional superposition
of massive objects (a $\sim500$nm radius nano-object) by introducing
a coupled atom-nanoparticle hybrid system and discussed how to detect
them. It will extend the demostration of the superposition principle
to unprecedented regimes of mass, $10^{8}$ times the current record~\citep{fein2019quantum}.
The method has several appealing features. It works for a generic
initial state, the control and readout of the motional state is through
well established versatile atomic protocols, and the created superposition
is very well protected from deleterious decoherence effects.

\prlsec{Acknowledgements} We acknowledge support from EPSRC grant
EP/N031105/1. MT acknowledges funding by the Leverhulme Trust (RPG-2020-197).

\appendix

\section{Atom-Nanoparticle motion and internal transitions}

We discuss the center-of-mass variables (Sec.~A), which allows to
reduce the problem to the effective interaction between the motional
state of the nanoparticle (Sec.~B) and the internal hyperfine state
of the atom (Sec.~C).

\subsection{Center-of-mass motion}

We introduce the center-of-mass (c.o.m.) variables 
\begin{alignat}{1}
\hat{R} & =\frac{m_{n}\hat{x}_{n}+m_{a}\hat{x}_{a}}{m_{n}+m_{a}},\qquad\hat{r}=\hat{x}_{n}-\hat{x}_{a},\label{eq:Rr}
\end{alignat}
where $\hat{R}$ ($\hat{r}$) is the c.o.m. (relative) position. The
corresponding zero-point motions are given by $\delta_{n}=\sqrt{\frac{\hbar}{2M\omega_{n}}}$
and $\delta_{a}=\sqrt{\frac{\hbar}{2\mu\omega_{a}}}$ , where we have
introduced the total mass $M=m_{n}+m_{a}\sim m_{n}$ and the reduced
mass $\mu=\frac{m_{a}m_{n}}{M}\sim m_{a}$. We define the mechanical
modes as 
\begin{equation}
\hat{R}=\delta_{n}(\hat{a}+\hat{a}^{\dagger})\qquad\hat{r}-d=\delta_{a}(\hat{b}+\hat{b}^{\dagger}),\label{eq:canonical}
\end{equation}
and using Eq.~(1) we readily find the nanoparticle-atom Hamiltonian:
\begin{equation}
H_{\text{nano-atom}}=\hbar\omega_{n}\hat{a}^{\dagger}\hat{a}+\hbar\omega_{a}\hat{b}^{\dagger}\hat{b}.\label{eq:nano-atom}
\end{equation}
We will be primarily interested in controlling the c.o.m. mode $\hat{a}$
which to good approximation coincides with the motion of the nanoparticle.
We consider the rigid-coupling regime discussed in the main text,
i.e. we prepare the atom in the motional ground state and require
$\delta_{n}\gg\delta_{a}$. More specifically, we require that the
displacement beam will not excite the atom's motional state, while
sufficiently exciting the nanoparticle.

Some remarks about the approximations involved are in order. In Eq.~(\ref{eq:nano-atom})
we have neglected terms of order $\sim\mathcal{O}(m_{a}/m_{n})$ which
for typical atomic and nanoscale masses would correspond to a correction
of $1$ part in $\sim10^{8}$. The analysis was also based on a semiclassical
approximation, where the internal motion responsible for the atomic
polarizability is assumed to reach a steady-state on a time-scale
faster than the motional time-scale of the atom in the trap~\cite{GRIMM200095}.
The full dynamics would require simultaneous integration of the optical
Bloch equations together with the atom-nanoparticle motional dynamics
as described by the quantum kinetic equations~\cite{balykin2000electromagnetic,leibfried2003quantum,chang2002density}.
In the following we will also consider additional lasers for controlling
the motional state of the atom; we will suppose that the atom remains
stably trapped for the duration of the experiment~\cite{garraway2000theory,jun2001stability}.

\subsection{Nanoparticle potential}

The potential of the nanoparticle in the Paul trap is given by 
\begin{equation}
\hat{H}_{\text{nano}}=\frac{m_{n}\omega_{n}^{2}}{2}\hat{x}_{n}^{2}+m_{n}g_{E}\hat{x}_{n}-F\hat{x}_{n},\label{eq:Hnano}
\end{equation}
where we have introduced the gravitational force $m_{n}g_{E}$ as
well as the radiation pressure force $F$ generated by the trapping
laser for the atom (see Fig.~1). 

We first trap the nanoparticle in a relatively stiff Paul trap $\omega_{n}=\omega_{1}$
with the radiation pressure force $F$ constrained by the requirement
of stable trapping in the Paul trap. The latter is controlled by light
intensity $I$ which also sets the atomic trap frequncy $\omega_{a}$
in Eq.~(2). Given the large mass of the nanoparticle in comparison
with the atom's mass we can have both a small radiation pressure force
$F\sim m_{n}g_{E}$ as well as a high trapping frequency $\omega_{a}$
for the atom \textendash{} the latter is required to introduce a handle
on the nanoparticle's motion. 

We then release the nanoparticle by (i) softening the Paul trap frequency
from $\omega_{n}=\omega_{1}$ to $\omega_{n}=\omega_{2}$ as well
as (ii) reducing the radiation pressure such that $F\ll m_{n}g_{E}$.
The net result is a change of equilibrium position and for a transient
period the nanoparticle is in free fall evolving according to the
potential

\begin{equation}
\hat{H}_{\text{nano}}\approx m_{n}g_{E}\hat{x}_{n}.
\end{equation}
In a nutshell, the idea is to suddenly release the nanoparticle from
the trap and use laser fields to create a spatial superposition exploiting
the atom-nanopaticle coupling. We effectively create a Mach-Zehnder
type interferometer for the nanoparticle: we exploit the Earth's gravitational
acceleration $\sim g_{E}$ to impart a phase difference on the spatial
parts of the superposition, which is then transferred to the internal
state and read out. 

\subsection{Two-photon stimulated Raman transitions}

We consider two types of interactions, one that (a) controls the internal
state without affecting the motional state, and one that (b) displaces
the motional state of the nanoparticle without changing the internal
one~\cite{wineland1998experimental}. 

In the former case (a) one links the ground and excited hyperfine
states, i.e. the states $\vert\uparrow\rangle$ and $\vert\downarrow\rangle$,
respectively, through a third hyperfine state $\vert3\rangle$ using
lasers of frequencies $\omega_{1}$ and $\omega_{2}$: on resonance
we would have $\vert\omega_{1}-\omega_{2}-\Delta_{3}\vert=\omega_{h}$
with $\Delta_{3}$ a suitably chosen detuning from the state $\vert3\rangle$.
Furthermore, we assume that the corresponding wave-vectors, $\bm{k}_{1}$
and $\bm{k}_{2}$, are such that their difference $\delta\bm{k}=\bm{k}_{1}-\bm{k}_{2}$
is parallel to the vertical $x$-axis with the projection denoted
by $\delta k$. Formally the interaction Hamiltonian is again given
by Eq.~(3), where $\eta=\delta k\delta_{n}$, and the coupling is
given by $\Omega_{\uparrow\downarrow}\equiv\frac{g_{\uparrow3}^{*}g_{3\downarrow}}{\Delta_{3}}$.
If we work at the carrier frequency, i.e. $\delta t=0$, the dominant
term in the Hamiltonian is insensitive to $\delta k$ and the motional
state remains unaffected, i.e. we only change the hyperfine state.
In the latter case (b) one instead stimulates the transitions $\vert\downarrow\rangle\rightarrow\vert3\rangle$
and $\vert3\rangle\rightarrow\vert\downarrow\rangle$, resulting in
a coupling $\Omega_{\downarrow\downarrow}\equiv\frac{g_{\downarrow3}^{*}g_{3\downarrow}}{\Delta_{3}}$.
Here we want to induce big displacements of the nanoparticle for which
large values of $\delta k$ are preferrable, e.g. $\delta k\sim\vert\bm{k}_{1}\vert$,$\vert\bm{k}_{2}\vert$.
The Hamiltonian is still the one in Eq.~(3) with the formal replacement
$\sigma_{+}\rightarrow\mathbb{I}$, where $\mathbb{I}$ is the identity
matrix: now the hyperfine state is unaffected and the motional state
changes, i.e. a displacement beam.

\section{Classical evolution\label{sec:Classical-evolution}}

We consider the motion of a point particle of mass $m$ in a harmonic
trap with frequency $\omega$ in the Earth's gravitational field.
In particular, the total Hamiltonian of the problem is given by

\begin{alignat}{1}
H_{1} & =\frac{p_{1}^{2}}{2m}+\frac{1}{2}m\omega^{2}x_{1}^{2}+mg_{E}x_{1},\label{eq:H1}
\end{alignat}
where $x_{1}$ ($p_{1}$) denote the position and momentum observable,
and $g_{E}$ is the gravitational acceleration. Here we will denote
the Earth's gravitational acceleration by $g_{E}$ while reserving
the symbol $g$ for the corresponding coupling which depends on $\omega_{n}$
(see Eq.~\ref{eq:gcoupling}). In Eq.~(\ref{eq:H1}) the subscript
$1$ labels the reference frame. We also introduce a shifted reference,
i.e. reference frame $2$, where the positions and momenta are given
by

\begin{equation}
x_{2}=x_{1}+\frac{g_{E}}{\omega^{2}},\qquad p_{2}=p_{1},\label{eq:xp12}
\end{equation}
and the Hamiltonian is 
\begin{equation}
H_{2}=\frac{p_{2}^{2}}{2m}+\frac{1}{2}m\omega^{2}x_{2}^{2}.\label{eq:H2}
\end{equation}

We are ultimately interested in the evolution described in reference
frame $1$, i.e. the evolution arising from Eq.~(\ref{eq:H1}). However,
as we will see when discussing the quantum case, it is instructive
to compare it to description in the shifted reference frame 2, i.e.
the evolution arising from Eq.~(\ref{eq:H2}). Specifically, in reference
2 we find the solution to be a simple harmonic motion:

\begin{alignat}{1}
x_{2} & =x_{2}(0)\text{cos}(\omega t)+\frac{p_{2}(0)}{m\omega}\text{sin}(\omega t),\label{eq:x2}\\
p_{2} & =-m\omega x_{2}(0)\text{sin}(\omega t)+p_{2}(0)\text{cos}(\omega t).\label{eq:p2}
\end{alignat}
Using Eq.~(\ref{eq:xp12}) we then immediately find the solution
in reference frame 1:

\begin{alignat}{1}
x_{1}= & x_{1}(0)\text{cos}(\omega t)+\frac{p_{1}(0)}{m\omega}\text{sin}(\omega t)\nonumber \\
 & +\frac{g_{E}}{\omega^{2}}(\text{cos}(\omega t)-1),\label{eq:x1}\\
p_{1}= & -m\omega x_{1}(0)\text{sin}(\omega t)+p_{1}(0)\text{cos}(\omega t)\nonumber \\
 & -m\omega\frac{g_{E}}{\omega^{2}}\text{sin}(\omega t).\label{eq:p1}
\end{alignat}
We now consider two different limits. We note that by taking the limit
$g_{E}\rightarrow0$ we recover simple harmonic motion, for example
the whole experiment, including the trap, is in free fall, i.e. we
recover Eqs.~(\ref{eq:x2}) and (\ref{eq:p2}) with the formal replacement
$x_{2}\rightarrow x_{1}$, $p_{2}\rightarrow p_{1}$. On the other
hand, in the limit $\omega\rightarrow0$, i.e. we switch off the trap,
we find:

\begin{alignat}{1}
x_{1} & =x_{1}(0)+\frac{p_{1}(0)}{m}t-\frac{g_{E}t^{2}}{2},\label{eq:x1ff}\\
p_{1} & =p_{1}(0)-mg_{E}t,\label{eq:p1ff}
\end{alignat}
as expected for free fall.

To relate the results to a quantum analysis we introduce the zero-point
motions, $\delta_{x}=\sqrt{\frac{\hbar}{2m\omega}}$ and $\delta_{p}=\sqrt{\frac{\hbar m\omega}{2}}$,
and the adimensional position and momentum,

\begin{alignat}{1}
X_{1} & =\frac{x_{1}}{\delta_{x}}=a+a^{*},\qquad P_{1}=\frac{p_{1}}{\delta_{p}}=i(a^{*}-a).\label{eq:X1x1P1p1}
\end{alignat}
The gravitational potential becomes

\begin{equation}
U=\hbar gX_{1},
\end{equation}
where the gravitational coupling is 
\begin{equation}
g=g_{E}\sqrt{\frac{m}{2\hbar\omega}}.\label{eq:gcoupling}
\end{equation}
The transition from harmonic to free fall motion depends on the strength
of the frequencies $\omega$ and $g$, which we now explore. We rewrite
Eqs.~(\ref{eq:x1}) and (\ref{eq:p1}) using Eqs.~(\ref{eq:X1x1P1p1}):

\begin{alignat}{1}
X_{1}= & X_{1}(0)\text{cos}(\omega t)+P_{1}(0)\text{sin}(\omega t),\nonumber \\
 & +2\frac{g}{\omega}(\text{cos}(\omega t)-1)\label{eq:X1}\\
P_{1}= & -X_{1}(0)\text{sin}(\omega t)+P_{1}(0)\text{cos}(\omega t)\nonumber \\
 & -2\frac{g}{\omega}\text{sin}(\omega t).\label{eq:P1}
\end{alignat}
Taking the limit $g\rightarrow0$ amounts to vanishing third terms
on the righthand side in Eqs.~(\ref{eq:X1}) and (\ref{eq:P1}),
which is the expected result as discussed above. On the other hand,
naively taking the limit $\omega\rightarrow0$ in Eqs.~(\ref{eq:X1})
and (\ref{eq:P1}) does not give the free fall evolution: the reason
is that these have been derived from Eqs.~(\ref{eq:X1}) and (\ref{eq:P1})
by diving/mupltipliying with $\delta_{x}$ and $\delta_{p}$ which
depend on the harmonic frequenciy $\omega$. A similar problem is
encountered also by using the modes

\begin{equation}
a_{1}=\frac{X_{1}+iP_{1}}{2},\qquad a_{1}^{*}=\frac{X_{1}-iP_{1}}{2}.\label{eq:amodes}
\end{equation}
Specifically, from Eqs.~(\ref{eq:X1}) and (\ref{eq:P1}) we find:

\begin{alignat}{1}
a_{1}= & a_{1}(0)e^{-i\omega t}+\frac{g}{\omega}(e^{-i\omega t}-1),\label{eq:a1}
\end{alignat}
where we are again confronted on how to consider the limiting free-fall
case.

The problem of taking the limit $\omega\rightarrow0$ can be avoided
by considering small adimensional expansion parameters, $gt$ and
$\omega t$ \textendash{} to study the free-fall case, we choose to
expand to quadratic order. Following the latter procedure we find
from Eq.~(\ref{eq:a1}):

\begin{alignat}{1}
a_{1}\approx & a_{1}(0)\left[1-i\omega t-\frac{1}{2}\omega^{2}t^{2})\right]+igt-\omega\frac{gt^{2}}{2}.\label{eq:a1a}
\end{alignat}
If we move back to the position-momentum description we find:

\begin{alignat}{1}
x_{1}= & x_{1}(0)+\frac{p_{1}(0)}{m}t+x_{1}(0)\frac{\omega^{2}t^{2}}{2}-\frac{g_{E}t^{2}}{2},\label{eq:x1a}\\
p_{1}= & p_{1}(0)-m\omega^{2}x_{1}(0)t+p_{1}(0)\frac{\omega^{2}t^{2}}{2}-mg_{E}t.\label{eq:p1a}
\end{alignat}
Eqs.~(\ref{eq:x1a}) and (\ref{eq:p1a}) have extra $\omega$-dependent
terms which were absent in the $\omega\rightarrow0$ limit (see Eqs.~(\ref{eq:x1ff})
and (\ref{eq:p1ff})). Unlike the former $\omega\rightarrow0$ calculation,
the approximation procedure is not state-independent, but depends
on the value of $x_{1}(0)$ and $p_{1}(0)$. In order to recover \emph{exactly
}free-fall one is implicitly assuming that the initial position and
momentum, $x_{1}(0)$ and $p_{1}(0)$, are small enough when taking
the $\omega\rightarrow0$ limit.

However, as we will explicitly see in the next sections we can retain
the additional $\omega$-dependent terms as they do not change the
induced gravitational phase \textendash{} as long as $\omega t$ remains
small. Furthermore, higher order harmonic terms \textendash{} beyond
the free fall approximation \textendash{} are interesting on its own
and could be used to ascertain the spatial superposition of large
nanoparticles without resorting to a dynamical equilibrium change
(see section~\ref{sec:Phase-difference}).

\section{Quantum evolution\label{subsec:Quantum-case-for}}

In this section we consider the quantum dynamics of a particle of
mass $m$ harmonically trapped and subject to the Earth's gravitational
potential. We continue to use the notation of Sec.~\ref{sec:Classical-evolution}
where the observables, e.g. $O$, are promoted to operators, e.g.
$O\rightarrow\hat{O}$. The classical analysis of the transition from
harmonic to free fall motion \textendash in particualr the approximations
involved \textendash{} carry over also to the quantum case. To simplify
the notation we will omit the subscript $1$ for quantities related
to reference frame $1$ most of the time.

\begin{figure}
\includegraphics[width=1\columnwidth]{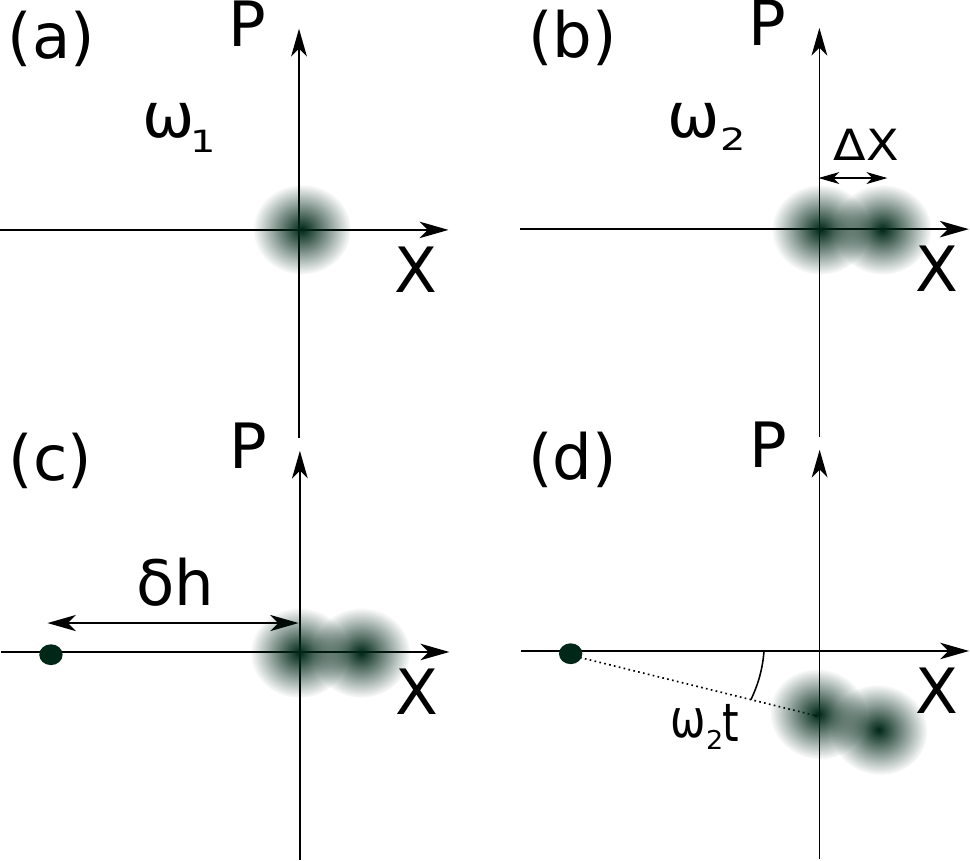}

\caption{We consider the \emph{vertical} motion of a particle in a Paul trap
in an Earth-bound laboratory. (a) The nanoparticle is initially confined
in a trap with frequency $\omega_{1}$ and kept close to the origin
of the trap; the gravitational force $mg_{E}$, where $m$ is the
mass of the nanoparticle and $g_{E}$ is the gravitational acceleration,
is counter-balanced by a radiation pressure force. (b) We change the
frequency to $\omega_{2}\ll\omega_{1}$ and create a small superposition
of size $\Delta x=\sqrt{\frac{\hbar}{2m\omega_{2}}}\Delta X.$ (c)
We decrease the radiation pressure force making it negligible with
respect to the gravitational one; this changes the equilibrium position
to $g_{E}/\omega_{2}^{2}=\sqrt{\frac{\hbar}{2m\omega_{2}}}\delta h$.
(d) We let the system evolve for a short time $t$ such that the motion
of the particle is governed by the uniform gravitational field. This
\emph{transient free fall} regime can be understood graphically \textendash{}
we note that the small arc drawn at radius $\delta h$ with subtended
angle $\omega_{2}t$ can be well approximated by the initial part
of a parabolic curve.}

\label{simulation-1-1} 
\end{figure}

\subsection{Change of equilibrium\label{subsec:Change-of-equilibrium}}

We consider the operator version of the Hamiltonian in Eqs.~(\ref{eq:H1})
which we rewrite as

\begin{alignat}{1}
\hat{H} & =\hbar\omega\hat{a}^{\dagger}\hat{a}+\hbar g(\hat{a}^{\dagger}+\hat{a}),\label{eq:H1q}
\end{alignat}
and an initial coherent state $\vert\alpha\rangle$ associated to
the $\hat{a}$ mode.

We first recall the definition of the displacement operator:

\begin{equation}
\hat{D}(\alpha)=e^{\alpha\hat{a}^{\dagger}-\alpha^{*}\hat{a}},\label{eq:D1}
\end{equation}
and the multiplication rule

\begin{equation}
\hat{D}(\alpha)\hat{D}(\beta)=e^{\frac{1}{2}(\alpha\beta^{*}-\alpha^{*}\beta)}\hat{D}(\alpha+\beta).\label{eq:D1rule}
\end{equation}
To find the time-evolution we restate the problem in a displaced frame:

\begin{alignat}{1}
\vert\alpha\rangle & \overset{\hat{D}}{\rightarrow}\vert\chi\rangle_{2}=\hat{D}(\delta)\vert\alpha\rangle,\label{eq:alpha2}\\
\hat{H} & \overset{\hat{D}}{\rightarrow}\hat{H}_{2}=\hat{D}(\delta)\hat{H}\hat{D}(\delta\alpha)^{\dagger},
\end{alignat}
where $\delta\equiv\frac{g}{\omega}$. In particular, we find $\hat{H}_{2}=\hbar\omega\hat{a}^{\dagger}\hat{a}$
and using Eqs.(\ref{eq:D1})-(\ref{eq:D1rule}) we find the time evolved
state

\begin{equation}
\vert\chi\rangle_{2}\rightarrow\vert\chi_{t}\rangle_{2}=e^{\frac{g}{2\omega}(\alpha^{*}-\alpha)}\vert(\alpha+\frac{g}{\omega})e^{-i\omega t}\rangle.
\end{equation}
We now go back to the original frame using the inverse transformation

\begin{alignat}{1}
\vert\chi_{t}\rangle_{2} & \overset{\hat{D}^{\dagger}}{\rightarrow}\hat{D}^{\dagger}(\delta)\vert\chi_{t}\rangle_{2}
\end{alignat}
Using again Eqs.~(\ref{eq:D1}) and (\ref{eq:D1rule}) we finally
find the time evolution of the state in the original frame:

\begin{alignat}{1}
\vert\alpha\rangle\rightarrow & e^{\frac{g}{2\omega}\left[\alpha^{*}(1-e^{i\omega t})-\alpha(1-e^{-i\omega t})\right]}\vert\alpha e^{-i\omega t}+\frac{g}{\omega}(e^{-i\omega t}-1)\rangle,
\end{alignat}
We expand to order $\mathcal{O}(t^{2})$ analogously as in the classical
case:

\begin{alignat}{1}
\vert\alpha\rangle\rightarrow & e^{-\frac{i}{2}(\alpha^{*}+\alpha)gt}e^{\frac{1}{2}(\alpha^{*}-\alpha)\omega\frac{gt^{2}}{2}}\nonumber \\
 & \vert\alpha(1-i\omega t-\frac{1}{2}\omega^{2}t^{2})-igt-\frac{\omega gt^{2}}{2}\rangle,\label{eq:result}
\end{alignat}
where we recognize in the first and second prefactors on the righthand
side a boost and a translation, respectively. In particular, using
Eq.~(\ref{eq:gcoupling}) the phase factors expressed become

\begin{alignat}{1}
-i\frac{(\alpha^{*}+\alpha)gt}{2} & =-i\frac{1}{2}\frac{x}{\hbar}\frac{g_{E}t}{2}\label{eq:ph1}\\
\frac{1}{2}(\alpha^{*}-\alpha)\omega\frac{gt^{2}}{2} & ==-i\frac{1}{2}\frac{p}{\hbar}\frac{g_{E}t^{2}}{2}\label{eq:ph2}
\end{alignat}
where $x=\delta_{x}(\alpha^{*}+\alpha)$ and $p=i\delta_{p}(\alpha^{*}-\alpha)$.
Similary, the state of the system $\vert\alpha\rangle$ has now been
been boosted by$-gt$ as well as displaced by $-\frac{\omega gt^{2}}{2}$
in accordance with the classical evolution in Eq.~(\ref{eq:a1a}).

\subsection{Change of equilibrium and frequency}

We consider the time-dependent Hamiltonian:

\begin{equation}
\hat{H}(t)=\frac{\hat{p}^{2}}{2m}+\frac{m\omega(t)^{2}}{2}\hat{x}^{2}+m\omega(t)^{2}d(t)\hat{x},\label{eq:H1f}
\end{equation}
where $\hat{x}$ and $\hat{p}$ are the operators associated to the
reference frame centered at the Paul-trap origin, i.e. reference frame
$1$. In particular, we have a sudden change of equilibrium position,
$d(t)$, and of the Paul trap frequency, $\omega(t)$, i.e.,

\begin{alignat}{1}
\omega(t) & =\begin{cases}
\omega_{1}, & t\leq0\\
\omega_{2}, & t>0
\end{cases},\label{eq:omegat}\\
d(t) & =\begin{cases}
0, & t\leq0\\
\frac{g_{E}}{\omega_{2}^{2}}, & t>0
\end{cases}.\label{eq:dt}
\end{alignat}
For $\omega_{2}=\omega_{1}$ one finds the problem already discussed
in the previous section~\ref{subsec:Change-of-equilibrium}.

Here we consider the full dynamics with the Hamiltonian defined in
Eqs.~(\ref{eq:H1f})-(\ref{eq:dt}). We consider an initial coherent
state $\vert\alpha\rangle$ associated to the mode $\hat{a}=\sqrt{\frac{\hbar}{2m\omega_{1}}}(\hat{x}+i\hat{p})$
prepeared at time $t=0$. The time-evolution for $t>0$ can be explicitly
computed~\cite{ma1989squeezing}:

\begin{equation}
\vert\alpha\rangle\rightarrow\hat{S}(z)\hat{D}(\epsilon)\hat{R}(\phi)\vert\alpha\rangle,\label{eq:alphasdr}
\end{equation}
where the operators are given by

\begin{alignat}{1}
\hat{S}(z) & =e^{\frac{1}{2}(z\hat{a}^{\dagger2}-z^{*}\hat{a}^{2})},\label{eq:Sz}\\
\hat{D}(\epsilon) & =e^{\frac{1}{2}(\epsilon\hat{a}^{\dagger}-\epsilon^{*}\hat{a})},\\
\hat{R}(\phi) & =e^{+i\phi\hat{a}^{\dagger}\hat{a}},
\end{alignat}
and the \emph{time-dependent }parameters are defined as follows 
\begin{alignat}{1}
e^{i\theta} & \tanh\vert z\vert=\frac{(e^{-2i\omega_{2}t}-1)\tanh r}{1-e^{-2i\omega_{2}t}\tanh^{2}r},\label{eq:zpar}\\
\epsilon & =\delta e^{i\phi}(1-e^{i\omega_{2}t})(\cosh r+e^{-i\omega_{2}t}\sinh r),\\
e^{i\phi} & =\frac{1-e^{2i\omega_{2}t}\tanh^{2}r}{\vert1-e^{2i\omega_{2}t}\tanh^{2}r\vert}e^{-i\omega_{2}t}.\label{eq:ephi}
\end{alignat}
We have two squeezing parameters: the customary one is given by $r=\frac{1}{2}\text{ln}(\frac{\omega_{2}}{\omega_{1}})$
and the dynamical one by $z=\vert z\vert e^{i\theta}$. The equilibirium
position in adimensional units is given by $\delta=\frac{g_{2}}{\omega_{2}}$
which is contained in the time-dependent parameter $\epsilon$, where
$g_{2}=g_{E}\sqrt{\frac{m}{2\hbar\omega_{2}}}$ is the coupling induced
by the gravitational acceleration.

We want to expand Eq.~(\ref{eq:alphasdr}) to order $\mathcal{O}(t^{2})$
during which the system is approximately in free fall as discussed
in the previous sections. However Eq.~(\ref{eq:alphasdr}) is not
yet in a suitable form as displacement and rotation operators preceed
the squeezing one; $\hat{S}(z)$ applied on a displaced coherent state
also changes its displacement. To avoid this problem we adapt the
analysis from~\cite{ma1989squeezing} to commute the operators:

\begin{equation}
\hat{S}(z)\hat{D}(\xi)=\hat{D}(\gamma)\hat{S}(z)\label{eq:sd}
\end{equation}
where 
\begin{alignat}{1}
\xi & =\epsilon+\alpha e^{i\phi},\\
\gamma & =\xi\cosh\vert z\vert-\xi^{*}\sinh\vert z\vert e^{i(\theta+\pi)}.
\end{alignat}
We can thus rewrite Eq.~(\ref{eq:alphasdr}) using Eq.~(\ref{eq:D1rule})
and Eq.~(\ref{eq:sd}) as

\begin{alignat}{1}
\vert\alpha\rangle\rightarrow & e^{\frac{1}{2}(\epsilon\alpha^{*}e^{-i\phi}+\epsilon^{*}\alpha e^{i\phi})}\hat{D}(\gamma)\hat{S}(z)\vert0\rangle\label{eq:alphasdr2}
\end{alignat}

We first note that the dynamical squeezing parameter $z$ in Eq.~(\ref{eq:zpar})
is only of order $\mathcal{O}(\omega_{1}t)$:

\begin{equation}
z=\frac{it\left(\text{\ensuremath{\omega_{1}^{2}}}-\omega_{2}^{2}\right)}{2\omega_{1}}\approx i\omega_{1}t.
\end{equation}
where we have assumed $\omega_{2}\ll\omega_{1}$. Wence we can neglect
squeezing and set $\hat{S}(z)\sim\mathbb{I}$ by assuming $\omega_{1}t\ll1$
(and hence also $\omega_{2}t\ll1$) . Performing a series expansion,
keeping only the relevant terms, we obtain from Eq.~(\ref{eq:alphasdr})
the following evolution:

\begin{alignat}{1}
\vert\alpha\rangle\rightarrow & e^{-\frac{i}{2}(\alpha^{*}+\alpha)\sqrt{\frac{\omega_{2}}{\omega_{1}}}g_{2}t}e^{\frac{1}{2}(\alpha^{*}-\alpha)\omega_{2}\sqrt{\frac{\omega_{1}}{\omega_{2}}}\frac{g_{2}t^{2}}{2}}\nonumber \\
 & \vert\alpha_{h}-i\sqrt{\frac{\omega_{2}}{\omega_{1}}}g_{2}t-\frac{\omega_{2}g_{2}t^{2}}{2}\sqrt{\frac{\omega_{1}}{\omega_{2}}})\rangle,\label{eq:alphasdr3}
\end{alignat}
where the harmonic contribution to the eigenvalue is given by

\begin{alignat}{1}
\alpha_{h}= & \alpha+\alpha(-i\frac{\omega_{1}^{2}+\omega_{2}^{2}}{2\omega_{1}}t-\frac{1}{2}\omega_{2}^{2}t^{2})\nonumber \\
 & +\alpha^{*}(i\frac{\omega_{1}^{2}-\omega_{2}^{2}}{2\omega_{1}}t+\frac{1}{4}\frac{\omega_{2}^{2}-\omega_{1}^{2}}{\omega_{1}^{2}}t^{2}).
\end{alignat}
It is instructive to introduce the gravitational coupling $g_{1}=g_{E}\sqrt{\frac{m}{2\hbar\omega_{1}}}$
associated to the modes $\hat{a}_{1}$, in particular, we note that
$g_{2}=\sqrt{\frac{\omega_{1}}{\omega_{2}}}g_{1}$. From (\ref{eq:alphasdr3})
then readily obtain the final result:

\begin{equation}
\vert\alpha\rangle\rightarrow e^{-\frac{i}{2}(\alpha^{*}+\alpha)g_{1}t}e^{\frac{1}{2}(\alpha^{*}-\alpha)\omega_{1}\frac{g_{1}t^{2}}{2}}\vert\alpha_{h}-ig_{1}t-\frac{\omega_{1}g_{1}t^{2}}{2}\rangle.\label{eq:alphasdr4}
\end{equation}
Relabelling $\omega1$ and $g_{1}$ as $\omega$ and $g$, respectively,
we recovered the result in Eq.~(\ref{eq:result}). In particular,
we note that the phase evolution depends only on $g_{E}$, but not
on the freequencies $\omega1$ or $\omega2$ \textendash{} see Eqs.~(\ref{eq:ph1})
and (\ref{eq:ph2}).

\section{Superposition state}

We consider the time evolution of the state $\vert\alpha\rangle$
and of the displaced state $\vert\alpha+\beta\rangle$ where $\beta\in\mathbb{R}$
according to Eq.~(\ref{eq:alphasdr4}). We readily find 
\begin{alignat}{1}
\vert\alpha\rangle & \rightarrow e^{i\xi}\vert\alpha'\rangle,\\
\vert\alpha+\beta\rangle & \rightarrow e^{-i\phi_{\text{grav}}}e^{i\xi}\vert\alpha'+\beta e^{i\phi}\rangle,
\end{alignat}
where $\xi=\frac{1}{2}(\alpha^{*}-\alpha)\omega\frac{gt^{2}}{2}$,
$\alpha'=\alpha_{h}-igt-\frac{\omega gt^{2}}{2}$, and the accumulated
phase difference is given by 
\begin{equation}
\phi_{\text{grav}}\equiv gt\beta.\label{eq:phigrav}
\end{equation}
By making the further approximation $\beta e^{i\phi}\approx\beta$
we recover the analysis from the main text \textendash{} the validity
of this approximation can be checked by evaluating Eq.~(\ref{eq:ephi}).
Note however that this latter assumption is not necessary and one
could still apply the protocol by modifying only step 7.

We now express the gravitational phase in terms of the physical quantities.
We first recall that $\beta=\Delta x/\delta_{R}$ where the zero-point
motion is $\delta_{R}=\sqrt{\frac{\hbar}{2m\omega}}$. Using Eq.~(\ref{eq:gcoupling})
we then readily recover Eq.~(4) from the main text,
i.e.,

\begin{equation}
\phi_{\text{grav}}=\frac{m_{n}g_{E}\Delta x\Delta t}{\hbar},\label{eq:phigravsupp}
\end{equation}
where we have set $t=\Delta t$. For a fixed $\Delta x$ this results
is indepedent of the Paul trap frequency as expected for the transient
free-fall motion.

On the other hand, the superposition size given by $\Delta x$ depends
on the Paul trap frequency $\omega_{n}$. In particular, applying
the displacement beam before or after we change the Paul trap frequency
from $\omega_{n}=\omega_{1}$ to $\omega_{n}=\omega_{2}$ can make
a big difference. This can be seen by recalling that $\Delta x=\delta_{R}\beta$
where $\delta_{R}=\sqrt{\frac{\hbar}{2m\omega_{n}}}$ is the zero-point
motion,$\beta=\Omega_{gg}\eta\delta t$ is the displacement generated
by the controlling lasers, and $\eta=k\delta_{R}$ is the Lamb-Dicke
parameter (see main text). In particular, combing the formulae we
readily find:

\begin{equation}
\Delta x=\frac{\hbar k}{2m\omega_{n}}\Omega_{gg}\delta t,
\end{equation}
where we explicitly see the $\sim\frac{1}{\omega_{n}}$ dependency
of the superposition size. In other words, applying the same displacement
beam in a weaker Paul trap leads to larger displacements as both the
zero-point motion $\delta_{R}$ and the Lamb-Dicke parameter $\eta$
contribute a factor $\frac{1}{\sqrt{\omega}}$.

The $\mathcal{O}(t^{3})$ correction to gravitational phase in Eq.~(\ref{eq:phigrav})
is given by 
\[
\phi^{(3)}=-\frac{1}{6}g\omega_{2}^{2}t^{3}\beta.
\]
If we require $\vert\phi^{(3)}\vert\ll\vert\phi_{\text{grav}}\vert$
we find the simple condition $\omega_{2}t\ll1$.

\section{Phase difference\label{sec:Phase-difference}}

It is instructive discusses the accumulated phase difference for spatial
superpositions in hamonic traps for long times. We have already discussed
the accumulation during the\emph{ transient free-fall motion} in case
there is a change of equilibrium position. We now ask what is the
accumulated phase difference when the motion can no longer be approximated
as free fall, for example, when the system undergoes \emph{a full
harmonic oscillation}. We perform this calculations using the semi-classical
approximation~\cite{storey1994feynman}.

\begin{figure}
\includegraphics[width=1\columnwidth]{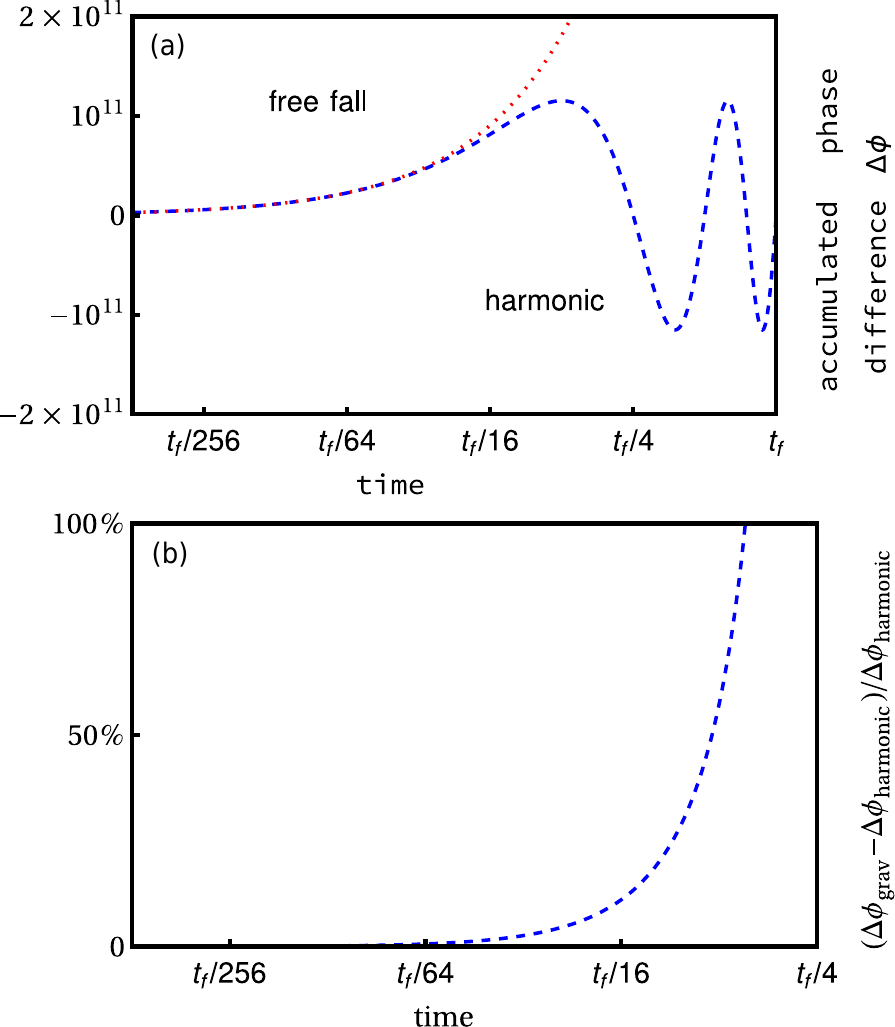}

\caption{(a) Accumulated phase difference $\Delta\phi$ for one oscillation
period $t_{f}=\frac{2\pi}{\omega}$. The blue dashed line corresponds
to $\Delta\phi_{\text{harmonic}}$ in Eq.~(\ref{eq:phiharmonic})
which oscillates at frequency $2\omega$ completing two full oscillations
in the trap oscillation period $t_{f}$. The red dotted line denotes
the transient free fall phase $\Delta\phi_{\text{grav}}$ in Eq.~(\ref{eq:harmonic2}).
We have considered typical values considered in the main text: the
nanoparticle mass $m=m_{n}\sim10^{-15}\text{m}$, Paul trap frequency
$\omega\sim5\times10^{-6}\text{Hz}$, initial position $x_{2}(0)=g_{e}/\omega^{2}\sim4\times10^{11}\text{m}$,
initial momentum to $p_{2}(0)\sim0,$ and superposition size $\Delta x=10^{-14}\text{m}$.
We find that one period of oscillation is $t_{f}\sim10^{6}\text{s}$.
(b) Relative error between the full harmonic solution and the
free fall approximation. The free-fall transient is a good approximation
for $t\apprle t_{f}/10\sim10^{5}\text{s}$, much longer than the time-scale
of the experiment.}

\label{transient} 
\end{figure}

Using the notation of section~(\ref{sec:Classical-evolution}) we
consider the description from reference frame 2, i.e. the dynamics
is purely harmonic with the Hamiltonian given in Eq.~(\ref{eq:H2}).
Here for simplicity we consider the case $\omega=\omega_{1}=\omega_{2}$.
The accumulated phase is given by the classical action

\begin{equation}
\phi[x_{2}(0),p_{2}(0)]=\frac{1}{\hbar}\int_{0}^{t}\left[\frac{p_{2}^{2}(s)}{2m}-\frac{m\omega^{2}}{2}x_{2}^{2}(s)\right]ds,
\end{equation}
where $x_{2}$ and $p_{2}$ are given in Eqs.~(\ref{eq:x2}) and
(\ref{eq:p2}). Evaluating the integral we readily find:

\begin{alignat}{1}
\phi[x_{2}(0),p_{2}(0)]= & \frac{\sin(2\omega t_{f})(p_{2}^{2}(0)-(m\omega(x_{2}(0))^{2})}{4m\omega\hbar}\nonumber \\
 & -\frac{p_{2}(0)x_{2}(0)}{\hbar}\sin^{2}(\omega t).\label{eq:phi}
\end{alignat}

We now consider the phase difference at different heights

\begin{equation}
\Delta\phi=-(\phi[x_{2}(0)+\Delta x,p_{2}(0)]-\phi[x_{2}(0),p_{2}(0)]),
\end{equation}
Using Eq.~(\ref{eq:phi}) we immediately find

\begin{alignat}{1}
\Delta\phi_{\text{harmonic}}= & \frac{\text{\ensuremath{\Delta x}}m\omega(\text{\ensuremath{\Delta x}}+2x_{2}(0))}{4\hbar}\sin(2\omega t)\nonumber \\
 & +\frac{\text{\ensuremath{\Delta x}}p_{2}(0)}{\hbar}\sin^{2}(\omega t)\label{eq:phiharmonic}
\end{alignat}
Let us expand the expression for small $\Delta x$ compared to $x_{2}(0)$
and to $\mathcal{O}(t)$, i.e. we are interested in the free-fall
regime of tiny superpositions. We readily find

\begin{alignat}{1}
\Delta\phi_{\text{grav}}\approx & \frac{\text{\ensuremath{\Delta x}}x_{2}(0)m\omega^{2}t}{\hbar}\label{eq:harmonic2}
\end{alignat}
Using $g_{e}=x_{2}(0)\omega^{2}$ we again recover Eq.~(\ref{eq:phigravsupp})
obtained from a more refined analysis. We have plotted in Fig.~\ref{transient}
a comparison between $\Delta\phi_{\text{harmonic}}$ and $\Delta\phi_{\text{grav}}$.

\bibliographystyle{unsrt}
\bibliography{nanoatom}

\end{document}